&pdflatex

\documentclass[12pt,prd,aps,amssymb,amsmath,tightenlines,showpacs]{revtex4}
\usepackage{graphicx}

\def\half{\textstyle{\frac{1}{2}}}
\def\threehalf{\textstyle{\frac{3}{2}}}
\def\cP{\mathcal P}
\def\cC{\mathcal C}
\def\cT{\mathcal T}
\def\veps{\varepsilon}

\begin{document}

\topmargin=0.0cm

\title{$\cP\cT$ phase transition in multidimensional quantum systems}

\author{Carl~M.~Bender${}^1$}
\email{cmb@wustl.edu}

\author{David~J.~Weir${}^2$}
\email{david.weir03@imperial.ac.uk}

\affiliation{${}^1$Department of Physics, Kings College London, Strand, London
WC2R 1LS, UK \footnote{Permanent address: Department of Physics, Washington
University, St. Louis, MO 63130, USA.} \\ ${}^2$Blackett Laboratory, Imperial
College, London SW7 2AZ, UK \footnote{Permanent address: Helsinki Institute
of Physics, P.O.~Box 64, University of Helsinki, 00014 Helsinki, Finland.}}

\begin{abstract}
Non-Hermitian $\cP\cT$-symmetric quantum-mechanical Hamiltonians generally
exhibit a phase transition that separates two parametric regions, (i) a region
of unbroken $\cP\cT$ symmetry in which the eigenvalues are all real, and (ii) a
region of broken $\cP\cT$ symmetry in which some of the eigenvalues are complex.
This transition has recently been observed experimentally in a variety of
physical systems. Until now, theoretical studies of the $\cP\cT$ phase
transition have generally been limited to one-dimensional models. Here, four
nontrivial coupled $\cP\cT$-symmetric Hamiltonians, $H=\half p^2+\half x^2+\half
q^2+\half y^2+igx^2y$, $H=\half p^2+\half x^2+\half q^2+y^2+igx^2y$, $H=\half
p^2+\half x^2+\half q^2+\half y^2+\half r^2+\half z^2+igxyz$, and $H=\half p^2+
\half x^2+\half q^2+y^2+\half r^2+\threehalf z^2+igxyz$ are examined. Based on
extensive numerical studies, this paper conjectures that all four models exhibit
a phase transition. The transitions are found to occur at $g\approx 0.1$, $g
\approx 0.04$, $g\approx 0.1$, and $g\approx 0.05$. These results suggest that
the $\cP\cT$ phase transition is a robust phenomenon not limited to systems
having one degree of freedom.
\end{abstract}

\maketitle

\section{Introduction}
\label{s1}

We believe that in order to advance the theory of $\cP\cT$ quantum mechanics it
is crucially important to answer the following natural question: Is there a $\cP
\cT$ phase transition in higher-dimensional quantum systems, or is $\cP\cT$
quantum mechanics just limited to one-dimensional models?

Recently, there have been some modest attempts to study what happens when two
$\cP\cT$-symmetric systems are coupled and when a $\cP\cT$-symmetric system is
coupled to a conventionally Hermitian system (see, for example, Ref.~\cite{R1}),
but in these studies only trivial matrix and harmonic-oscillator models were
considered. In early analytical approaches (see, for example,
Refs.~\cite{R2,R3}) some progress was made in calculating the $\cC$ operator for
some complicated $\cP\cT$-symmetric Hamiltonians; showing that the $\cC$
operator exists is equivalent to proving that the eigenvalues are real
\cite{R5}. However, calculating the $\cC$ operator is difficult, and $\cC$ was
only calculated to first order in perturbation theory in Refs.~\cite{R2,R3}.

In this paper we report a direct numerical attack on some nontrivial coupled
Hamiltonians that were first considered in Ref.~\cite{R3}. Specifically, we
consider the four nontrivial Hamiltonians
\begin{equation}
H=\half p^2+\half x^2+\half q^2+\half y^2+igx^2y,
\label{E1}
\end{equation}
\begin{equation}
H=\half p^2+\half x^2+\half q^2+y^2+igx^2y,
\label{E2}
\end{equation}
\begin{equation}
H=\half p^2+\half x^2+\half q^2+\half y^2+\half r^2+\half z^2+igxyz,
\label{E3}
\end{equation}
\begin{equation}
H=\half p^2+\half x^2+\half q^2+y^2+\half r^2+\threehalf z^2+igxyz,
\label{E4}
\end{equation}
where the coupling constants $g$ are real parameters. For these Hamiltonians a
convincing demonstration that there exists a range of $g$ for which the
eigenvalues are all real requires the accurate calculation of thousands of
eigenvalues. Initial nondefinitive investigations using comparatively elementary
numerical methods suggested that there might be complex eigenvalues for all
nonzero values of $g$ \cite{R6}. If this were true, then we would be forced to
admit that $\cP\cT$ quantum mechanics is theoretically interesting but of
limited scope.

This paper presents numerical evidence that for each of these complex
Hamiltonians there are actually ranges of $g$ for which the eigenvalues of these
Hamiltonians are all real. Based on this numerical work, we conjecture that for
$H$ in (\ref{E1}) the critical value of $g$ is about $0.1$ and that when $|g|<
0.1$ the eigenvalues are {\it all} real. For the Hamiltonians in (\ref{E2}),
(\ref{E3}), and (\ref{E4}) the critical values of $g$ are about $0.04$, $0.1$,
and $0.05$, and when $|g|$ is less than these values, the eigenvalues are all
real. To obtain these results we have used a powerful numerical scheme known as
the {\it implicitly restarted Arnoldi method} \cite{LS}. We have calculated many
thousands of eigenvalues accurate to about one part in $10^6$ and have used
about $20,000$ hours of CPU time. Our results suggest that $\cP\cT$ quantum
mechanics is general and robust, and that it extends to genuinely nontrivial
higher-dimensional quantum-mechanical models. At present we can only draw
conclusions based on detailed numerical studies, but the results in this paper
suggest that $\cP\cT$ symmetry may even extend to infinite-dimensional
quantum-field-theoretic models.

This paper is organized simply. In Sec.~\ref{s2} we describe the $\cP\cT$ phase
transition and in Sec.~\ref{s3} we summarize our numerical approach and present
our results in graphical form. Section \ref{s4} contains brief concluding
remarks.

\section{$\cP\cT$ Phase Transition}
\label{s2}

A Hamiltonian is $\cP\cT$ symmetric if it is invariant under combined space
reflection (parity) $\cP$ and time reversal $\cT$. Such a Hamiltonian is said to
have an {\it unbroken} $\cP\cT$ symmetry if its eigenfunctions are also
eigenstates of the $\cP\cT$ operator. When the $\cP\cT$ symmetry of a
Hamiltonian is unbroken, its eigenvalues are all real even though the
Hamiltonian may not be Dirac Hermitian \cite{R7}. (We use the term {\it Dirac
Hermitian} to describe a linear operator that remains invariant under the
combined operations of matrix transposition and complex conjugation.) Such a
$\cP\cT$-symmetric Hamiltonian is physically relevant because it generates
unitary time evolution \cite{R5}.

$\cP\cT$-symmetric Hamiltonians usually depend on one or more parameters. The
Hamiltonians that have been studied so far typically possess an
unbroken-$\cP\cT$-symmetric phase (a parametric region of unbroken $\cP\cT$
symmetry in which all of the eigenvalues are real) and a broken-$\cP
\cT$-symmetric phase (a parametric region of broken $\cP\cT$ symmetry in which
some of the eigenvalues are complex). The boundary between these two regions is
the $\cP\cT$ phase transition and this transition occurs at a critical value of
a parameter. 

A simple example of a $\cP\cT$-symmetric Hamiltonian that has a $\cP\cT$ phase
transition is the $2\times2$ matrix Hamiltonian \cite{R8}
\begin{equation}
H=\left(\begin{array}{cc} re^{i\theta} & s \cr s & re^{-i\theta}
\end{array}\right),
\label{E5}
\end{equation}
where the three parameters $r$, $s$, and $\theta$ are real. This Hamiltonian is
not Dirac Hermitian but it is ${\cal PT}$ invariant, where the parity operator
$\cP$ is
\begin{equation}
\mathcal{P}=\left(\begin{array}{cc} 0 & 1 \cr 1 & 0 \end{array}\right)
\label{E6}
\end{equation}
and the time-reversal operator ${\cal T}$ is complex conjugation. The region of 
unbroken $\cP\cT$ symmetry is $s^2\geq r^2\sin^2\theta$.

Almost all studies of non-Hermitian $\cP\cT$-symmetric quantum-mechanical
Hamiltonians have focused on finite-dimensional matrix Hamiltonians like that in
(\ref{E5}) or on one-degree-of-freedom Hamiltonians of the type $H=p^2+V(x)$,
for which the condition of $\cP\cT$ symmetry is $V^*(x)=V(-x)$. The
Schr\"odinger eigenvalue problem $H\psi=E\psi$ for such Hamiltonians takes the
form of the ordinary differential equation $-\psi''(x)+V(x)\psi(x)=E\psi(x)$.

\begin{figure}[h!]
\begin{center}
\includegraphics[scale=0.32, viewport=0 0 842 595]{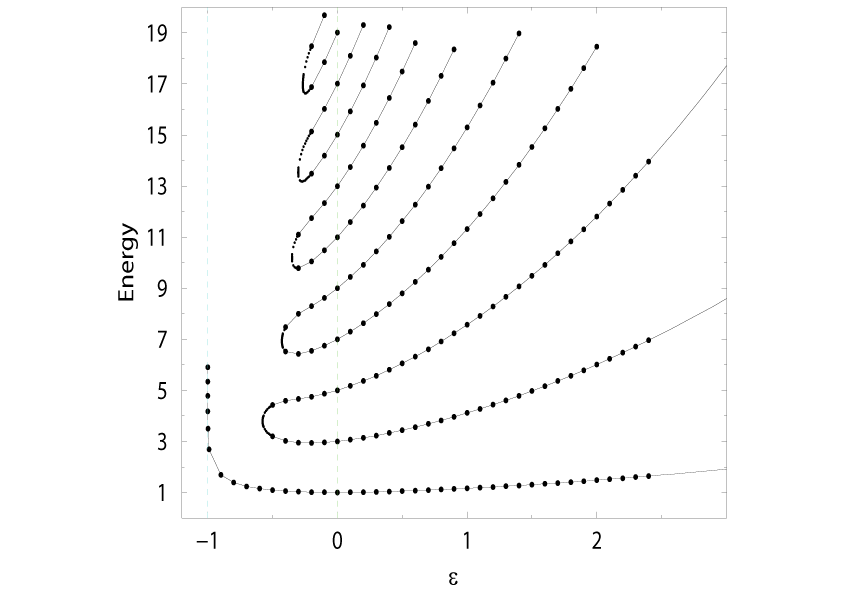}
\end{center}
\caption{Real eigenvalues of the $\cP\cT$-symmetric Hamiltonian (\ref{E7})
plotted as functions of $\veps$. The region of unbroken $\cP\cT$ symmetry is
$\veps\geq0$. In the region $\veps<0$ of broken $\cP\cT$ symmetry, the
eigenvalues emerge from the complex plane at exceptional points as degenerate
pairs; these pairs then split into real eigenvalues. As $\veps$ approaches the
critical point at $0$ from below, this decomplexification process begins
with the low-lying eigenvalues and terminates at $\veps=0$ at the highest-lying
eigenvalues.}
\label{F1}
\end{figure}

The first $\cP\cT$-symmetric Hamiltonian that was studied in detail has the form
\begin{equation}
H=p^2+x^2(ix)^\veps,
\label{E7}
\end{equation}
where $\veps$ is a real parameter \cite{R9,R10}. Dorey, Dunning, and Tateo
proved that the eigenvalues of the corresponding Schr\"odinger eigenvalue
equation
\begin{equation}
-\psi''(x)+x^2(ix)^\veps\psi(x)=E\psi(x)
\label{E8}
\end{equation}
are all real when $\veps\geq0$ \cite{R11,R12}. These eigenvalues are plotted as
functions of $\veps$ in Fig.~\ref{F1}. The critical value of $\veps$ is 0, and
the region of broken $\cP\cT$ symmetry is $\veps<0$. As $\veps$ approaches $0$
through negative values, complex-conjugate pairs of eigenvalues become
degenerate as they emerge from the complex plane and then split into real pairs
of eigenvalues. The values of $\veps$ at which the eigenvalues become degenerate
are sometimes called {\it exceptional points} \cite{RRR}. Observe from
Fig.~\ref{F1} that $\veps=0$ is the {\it limit point} of a sequence of
exceptional points. For this model the decomplexification process is a
high-energy phenomenon; that is, as $\veps$ approaches $0$ from below,
sequentially higher eigenvalues (rather than lower eigenvalues) become
degenerate. This is because $\veps$ is a {\it singular} perturbation parameter
\cite{R13}. (The same kind of limiting process was discovered many years ago for
the case of the anharmonic oscillator $H=p^2+x^2+gx^4$, except that for the
anharmonic oscillator the exceptional points lie in the complex-$g$ plane rather
than on the real-$g$ axis \cite{R13,R14}.)

Many other $\cP\cT$-symmetric Hamiltonians that exhibit $\cP\cT$ phase
transitions have been studied. For example, the Hamiltonian
\begin{equation}
H=p^2+x^4+iAx,
\label{E9}
\end{equation}
where $A$ is a real parameter, exhibits a $\cP\cT$ phase transition at $A=\pm
3.169$ \cite{R15}. In contrast to the Hamiltonian (\ref{E7}), as $A$ approaches
the phase transition through the region $|A|>3.169$ of broken $\cP\cT$ symmetry,
the eigenvalues emerge from the complex domain starting with the highest-lying
eigenvalues and ending with the lowest-lying eigenvalues. The exceptional points
remain well separated and do not converge to a limit point because $A$ is a {\it
regular} perturbation parameter \cite{R13}.

Another example of a $\cP\cT$-symmetric Hamiltonian having a $\cP\cT$ phase
transition is
\begin{equation}
H=-\frac{d^2}{d\theta^2}+ig\cos\theta,
\label{E10}
\end{equation}
where $g$ is a real parameter \cite{R16}. Here, the Schr\"odinger eigenvalue
problem is posed on a finite rather than on an infinite domain. If the
eigenfunctions are required to be $2\pi$ periodic and odd in $\theta$, the
region of unbroken $\cP\cT$ symmetry is $|g|<3.4645$ \cite{R16}.

A physically motivated example of a $\cP\cT$-symmetric Hamiltonian that exhibits
a $\cP\cT$ phase transition was discussed by Rubinstein, Sternberg, and Ma
\cite{R17}. Their Hamiltonian arises in the context of superconducting wires.
Again, the Schr\"odinger eigenvalue problem
\begin{equation}
-\psi''(x)-igx\psi(x)=E\psi(x)
\label{E11}
\end{equation}
is posed on the finite domain $|x|\leq 1$ and the boundary conditions are $\psi(
\pm1)=0$. The region of unbroken $\cP\cT$ symmetry is $g<12.31$, and the
behavior of the eigenvalues is qualitatively similar to that of $H$ in
(\ref{E10}). (Compare Fig.~1 of Ref.~\cite{R16} with Fig.~1 of Ref.~\cite{R17}.)

The $\cP\cT$ phase transition has been seen repeatedly in laboratory
experiments. It was first observed in optical wave guides
\cite{R18,R19,R20,R21,AAA}, but it has also been observed in atomic diffusion
\cite{R22}, lasers \cite{R23,R24}, superconducting wires \cite{R17}, nuclear
magnetic resonance \cite{R25}, and most recently in microwave cavities
\cite{R26} and in electronic circuits \cite{BBB}. The $\cP\cT$ phase
transition is a clear and prominent effect and not a subtle phenomenon;
laboratory measurements and theoretical predictions have agreed with virtually
no error.

Theoretical studies of eigenvalues (as in Fig.~\ref{F1}) and experimental
observations suggest that the $\cP\cT$ phase transition is quite generic, but
almost all of the theoretical and experimental work that has been done so far
has concerned systems that have just one degree of freedom. Indeed, the proof of
the reality of the eigenvalues in Refs.~\cite{R11,R12} relies on establishing a
correspondence, known as the {\it ODE/IM correspondence}, between ordinary
differential equations and integrable models. We emphasize that this
correspondence involves {\it ordinary} and not {\it partial} differential
equations. Thus, we are motivated to study in Sec.~\ref{s3} the eigenvalues of
the four multidimensional Hamiltonians in (\ref{E1}) - (\ref{E4}).

\section{Numerical calculation of eigenvalues}
\label{s3}

To find the eigenvalues of the Hamiltonians in (\ref{E1}) - (\ref{E4}), we
follow a calculational approach that was used in Ref.~\cite{R10}. We express the
Hamiltonian in harmonic-oscillator-basis states [see Eq.~(3.2) in
Ref.~\cite{R10}], truncate the Hamiltonian matrix to an $N\times N$ numerical
array, and then calculate the eigenvalues. In this paper the eigenvalues are
calculated using the implicitly restarted Arnoldi method as implemented by
ARPACK \cite{LSY}. This numerical technique works particularly well because the
array is {\it sparse}; that is, only a small percentage of the matrix elements
are nonzero. Furthermore, the matrix becomes increasingly sparse with increasing
dimension.

\begin{figure}[t!]
\begin{center}
\includegraphics[trim=10mm 17mm 53mm 28mm,clip=true,scale=0.50]{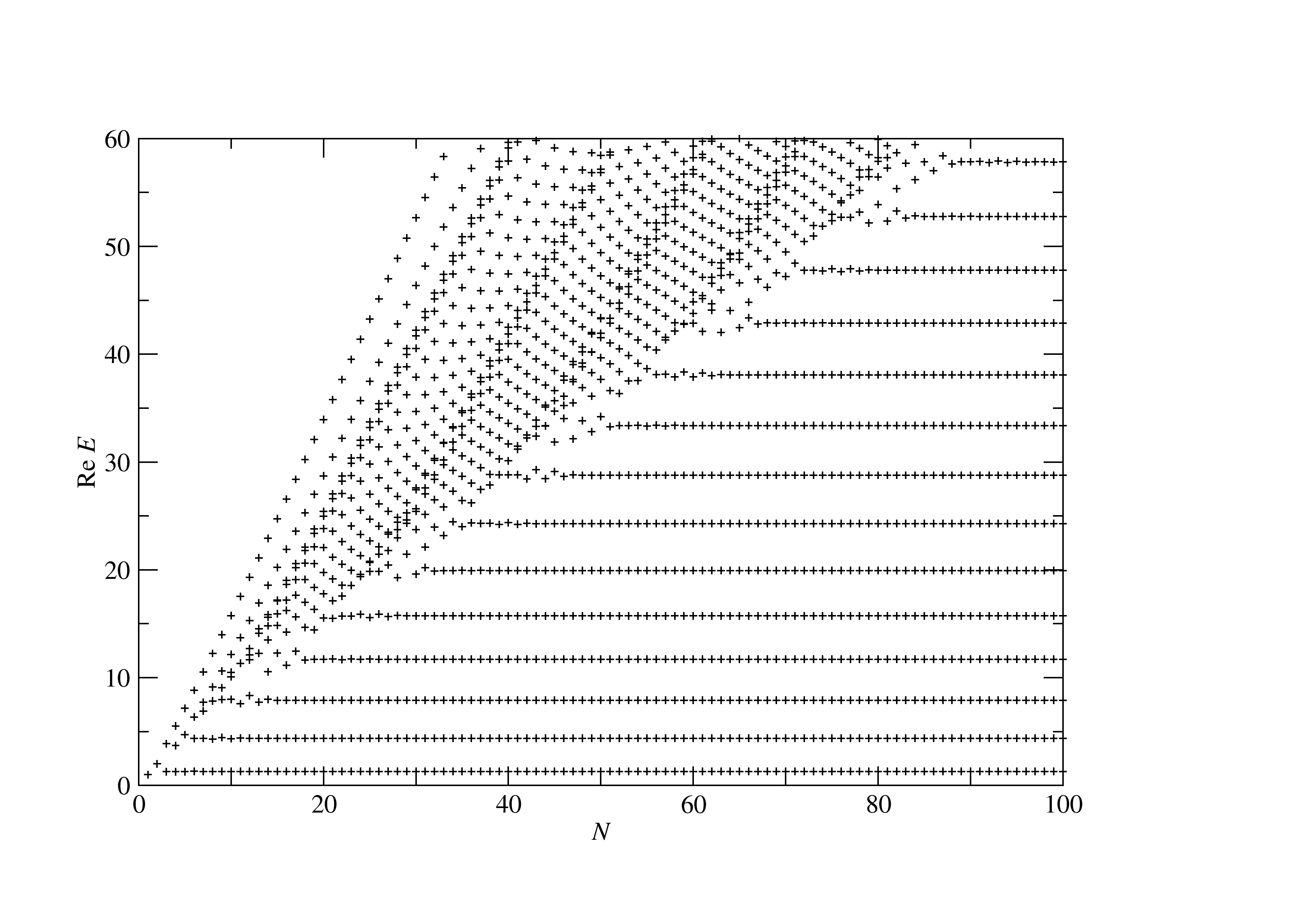}
\end{center}
\caption{Numerical convergence of the first seven eigenvalues of the Hamiltonian
in (\ref{E12}) using the implicitly restarted Arnoldi method. At first, the
eigenvalues vary chaotically as the dimension $N$ of the matrix increases, but
eventually, starting with the lowest energies, they settle down to their correct
numerical values as shown in Fig.~\ref{F1} at $\veps=1$.}
\label{F2}
\end{figure}

As is shown in Fig.~12 of Ref.~\cite{R10}, the convergence of the eigenvalues as
$N$ increases is noisy at first, but starting with the lowest energies, the
eigenvalues settle down one-by-one to their correct limiting values. In
Ref.~\cite{R10} the eigenvalues of the Hamiltonian $H=p^2+ix^3$ are calculated
as the dimension of the matrix ranges up to $N=15$. Here, to illustrate the
convergence of the eigenvalues, we plot in Fig.~\ref{F2} the eigenvalues of
\begin{equation}
H=p^2+x^2+ix^3
\label{E12}
\end{equation}
for $N$ ranging from 2 to 100. Note that the eigenvalues are converging to the
values shown in Fig.~\ref{F1} for $\veps=1$.

\subsection{Calculation of the Eigenvalues of the Hamiltonian (\ref{E1})}
\label{ss3a}

\begin{figure}[t!]
\begin{center}
\includegraphics[trim=10mm 17mm 53mm 28mm,clip=true,scale=0.50]{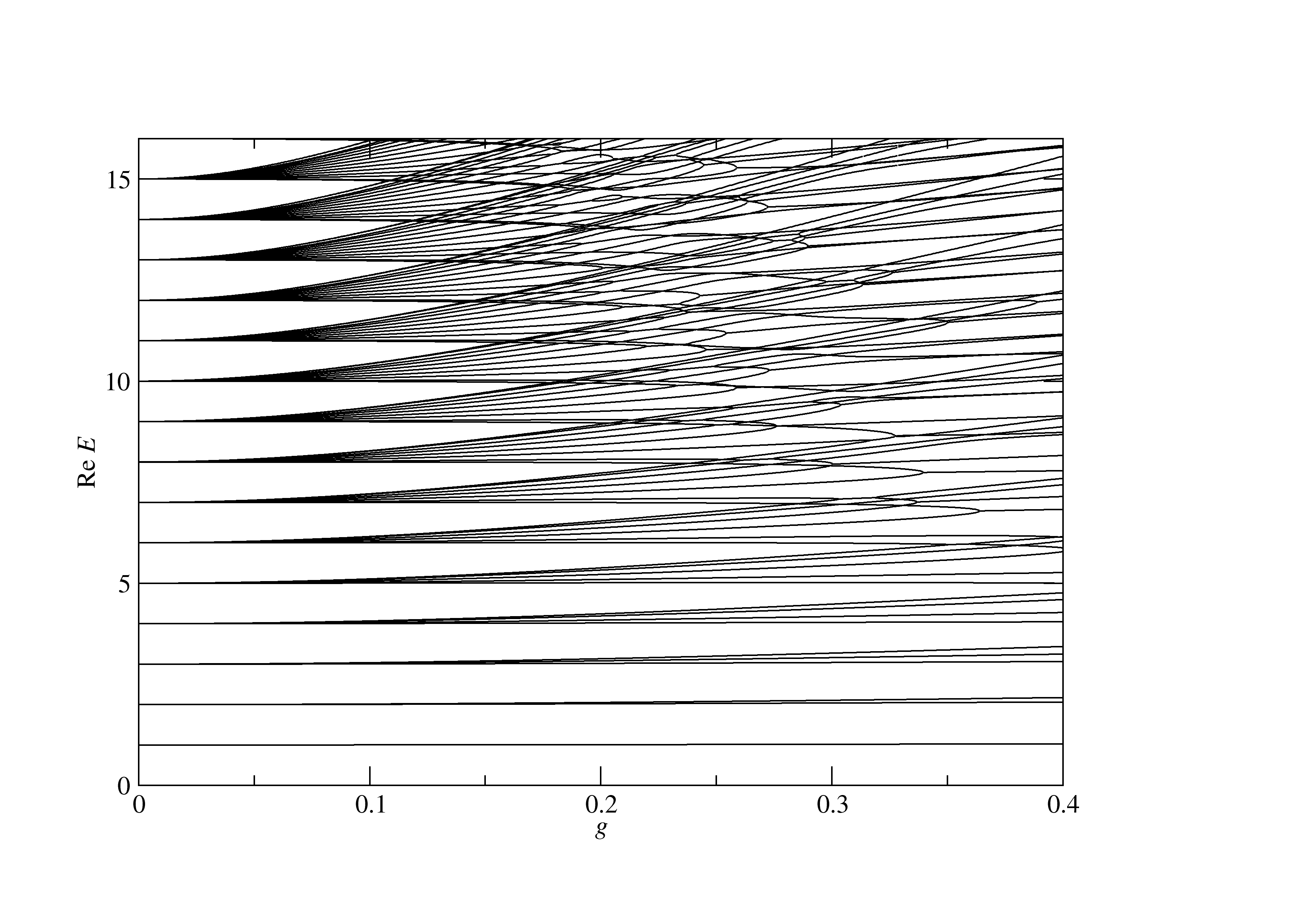}
\end{center}
\caption{Real parts of the eigenvalues of $H$ in (\ref{E1}) that have converged
for a $100^2\times 100^2$ matrix plotted versus $g$. The coupling constant $g$
ranges from $0$ to $0.4$ in steps of $0.0005$, and the eigenvalues shown have
real parts less than $16$. The plot is complicated because when $g=0$ the
eigenvalues become increasingly degenerate with increasing energy; there is one
eigenvalue of energy $1$, two eigenvalues of energy $2$, three eigenvalues of
energy $3$, and so on. As $g$ increases, the degeneracy is broken. Eventually,
one can see energy levels crossing, but most crossings only represent accidental
degeneracies. One must look carefully to see the exceptional points, which are
not so easy to see as those in Fig.~\ref{F1}. In the range of $g$ shown, the
lowest-energy exceptional point occurs at about $g=0.364$, where one of the
sixth and one of the seventh eigenvalues become degenerate and form a
complex-conjugate pair.}
\label{F3}
\end{figure}

\begin{figure}[t!]
\begin{center}
\includegraphics[trim=10mm 17mm 53mm 28mm,clip=true,scale=0.50]{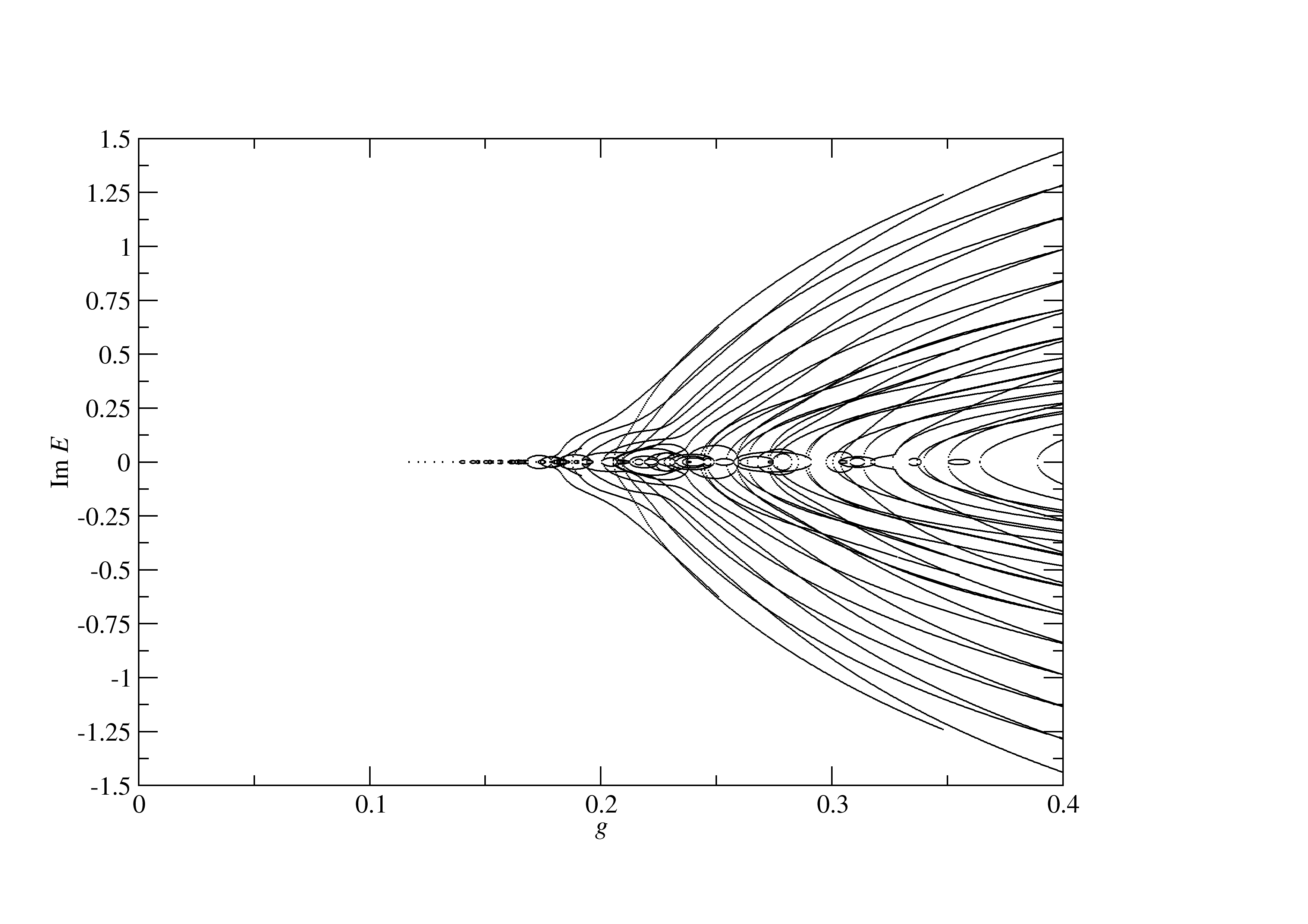}
\end{center}
\caption{Imaginary parts of the eigenvalues whose real parts are shown in
Fig.~\ref{F3}. Note that the imaginary parts of the eigenvalues are much
smaller than the real parts by roughly a factor of 10. For the size of the
matrix studied, it appears that the critical point is at $g\approx 0.1$. Below
this point there are no complex eigenvalues whose real parts are less than 16.}
\label{F4}
\end{figure}

If we use the same approach for $H$ in (\ref{E1}), the numerical convergence is
{\it faster} than that for $H$ in (\ref{E12}) because the matrix is more sparse;
only about $5\%$ of the matrix elements are nonzero. Using $100^2\times 100^2$
matrices ($10^8$ matrix elements), we calculate the eigenvalues for $g$ ranging
from $0$ to $0.4$ in steps of $0.0005$. There are $10^4$ eigenvalues, but we
limit our attention to those eigenvalues that have settled down and are changing
by less than one part in $10^6$ as we increase the size of the matrix from $80^2
\times 80^2$ to $90^2\times 90^2$ to $100^2\times 100^2$. In Figs.~\ref{F3} and
\ref{F4} we plot the real and imaginary parts of those eigenvalues whose real
parts range from $0$ to $16$. A key result is given in Fig.~\ref{F4}; it appears
that for $g$ greater than about $0.1$ there are many complex eigenvalues, but
that there are no complex eigenvalues when $g<0.1$. This suggests that there is
a critical value of $g$ near $0.1$.

\begin{figure}[t!]
\begin{center}
\includegraphics[trim=10mm 17mm 53mm 28mm,clip=true,scale=0.50]{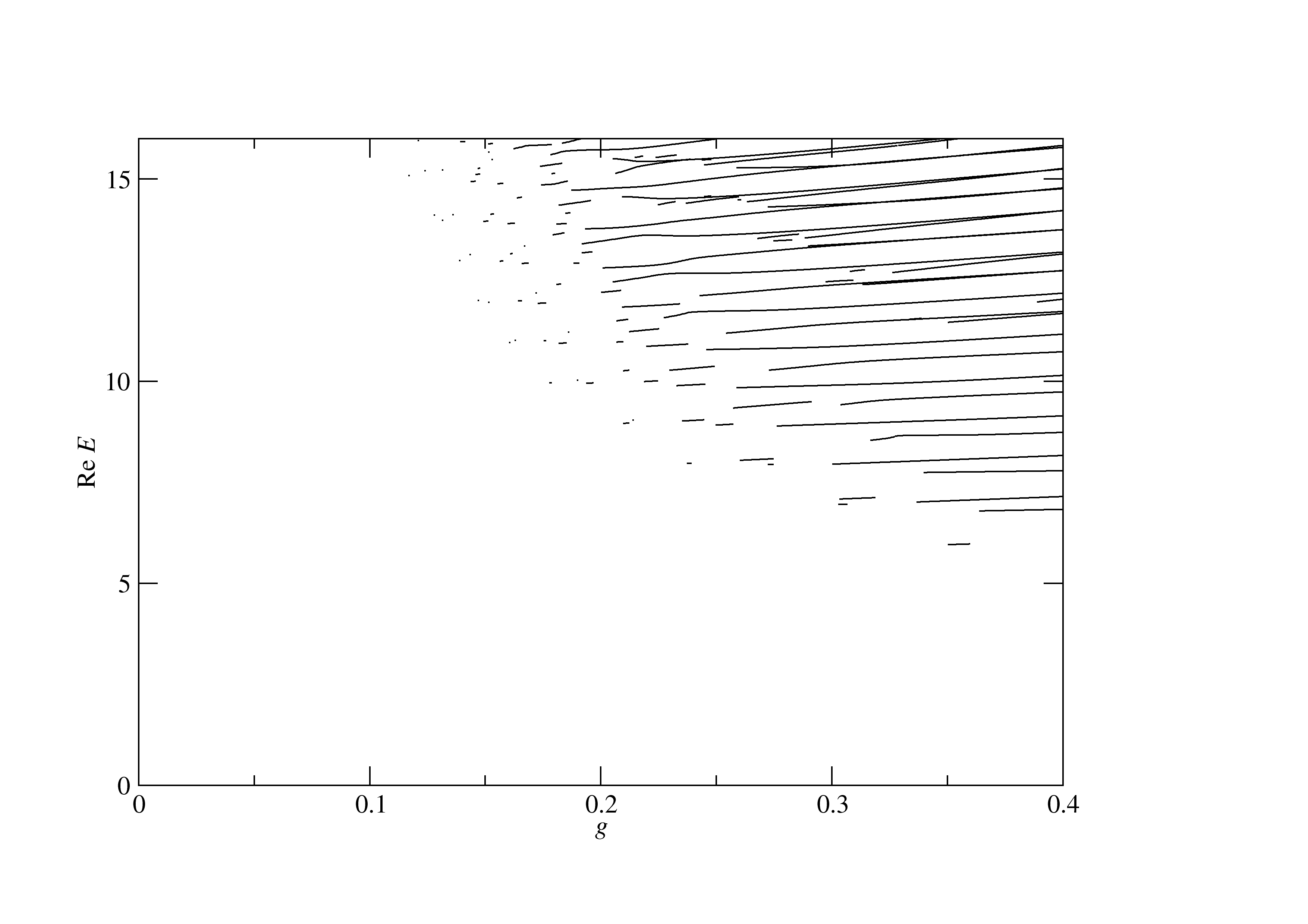}
\end{center}
\caption{Real parts of the eigenvalues in Fig.~\ref{F4} whose imaginary parts
are nonzero. Note that the real parts of these eigenvalues grow with decreasing
$g$. We can obtain a more accurate estimate of the critical value of $g$ by
tracing a curve through the left-most points on the graph.}
\label{F5}
\end{figure}

From the eigenvalues in Fig.~\ref{F4} we then select out just the eigenvalues
that have nonzero imaginary parts. The real parts of these eigenvalues are shown
in Fig.~\ref{F5}. Note that the real parts of these eigenvalues increase rapidly
as $g$ decreases. Thus, if there is a nonzero critical value of $g$ at which a 
$\cP\cT$ phase transition occurs, this transition is a high-energy phenomenon,
as opposed to the $\cP\cT$ phase transition of $H$ in (\ref{E7}) (see
Fig.~\ref{F1}).

If we trace a curve through the left-most points in Fig.~\ref{F5}, we can see
that this curve rises steeply as $g$ decreases, and we believe that this curve
becomes infinite at approximately $g\approx 0.1$. Unfortunately, it is not easy
to fit a curve numerically through these points because they are not very
regular. However, when we repeat these numerical calculations for the
Hamiltonian (\ref{E2}), we find that the corresponding points are more regular,
and it is indeed possible to fit such a curve.

\subsection{Calculation of the Eigenvalues of the Hamiltonian (\ref{E2})}
\label{ss3b}

\begin{figure}[t!]
\begin{center}
\includegraphics[trim=10mm 17mm 53mm 28mm,clip=true,scale=0.50]{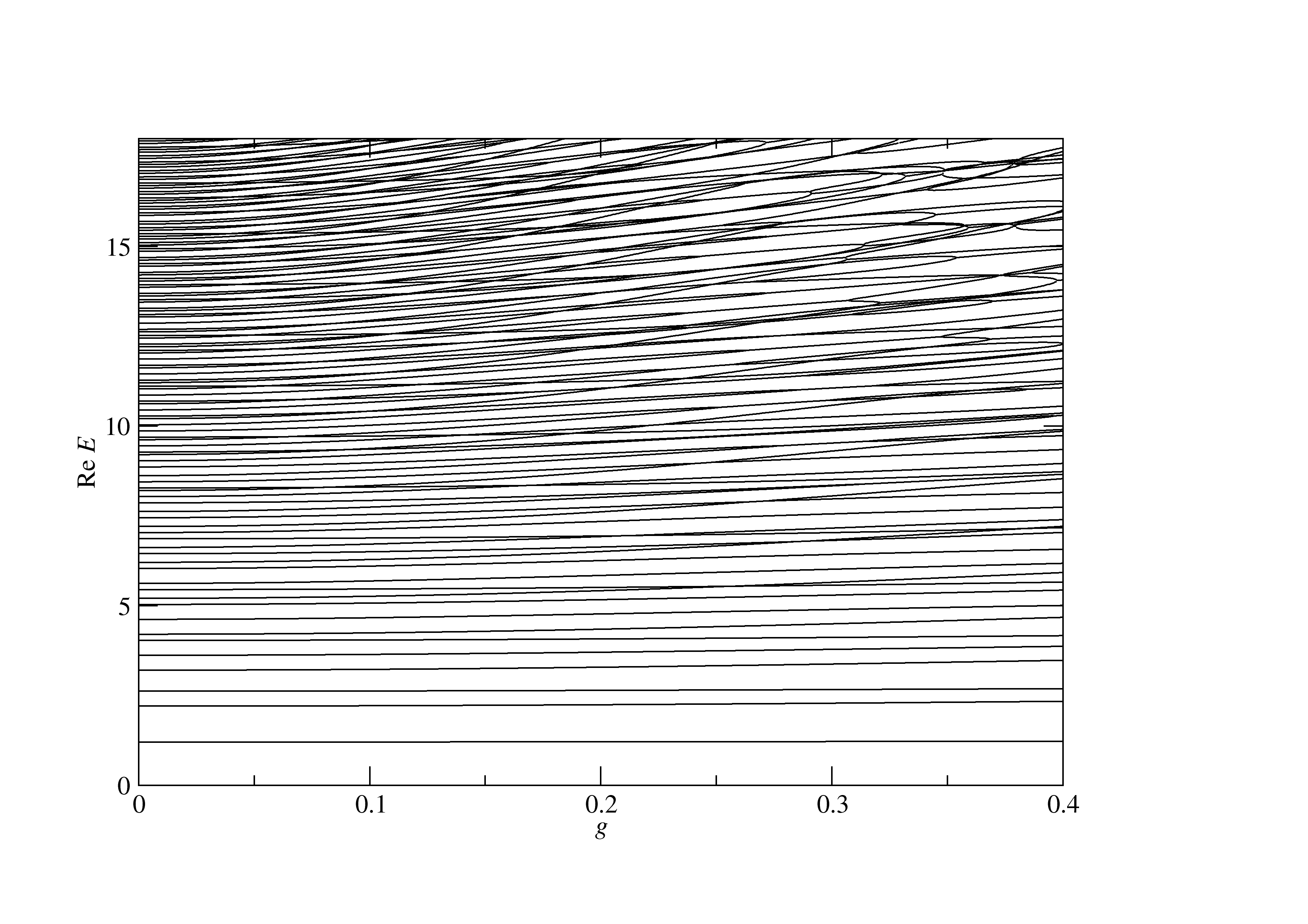}
\end{center}
\caption{Real parts of the eigenvalues of the $\cP\cT$-symmetric Hamiltonian
(\ref{E2}) plotted as functions of $g$ for $0\leq g\leq 0.4$ in steps of
$0.0005$. All eigenvalues (both real and complex) whose real parts are less than
18 are shown. This graph is the analog of Fig.~\ref{F3}.}
\label{F6}
\end{figure}

\begin{figure}[t!]
\begin{center}
\includegraphics[trim=10mm 17mm 53mm 28mm,clip=true,scale=0.50]{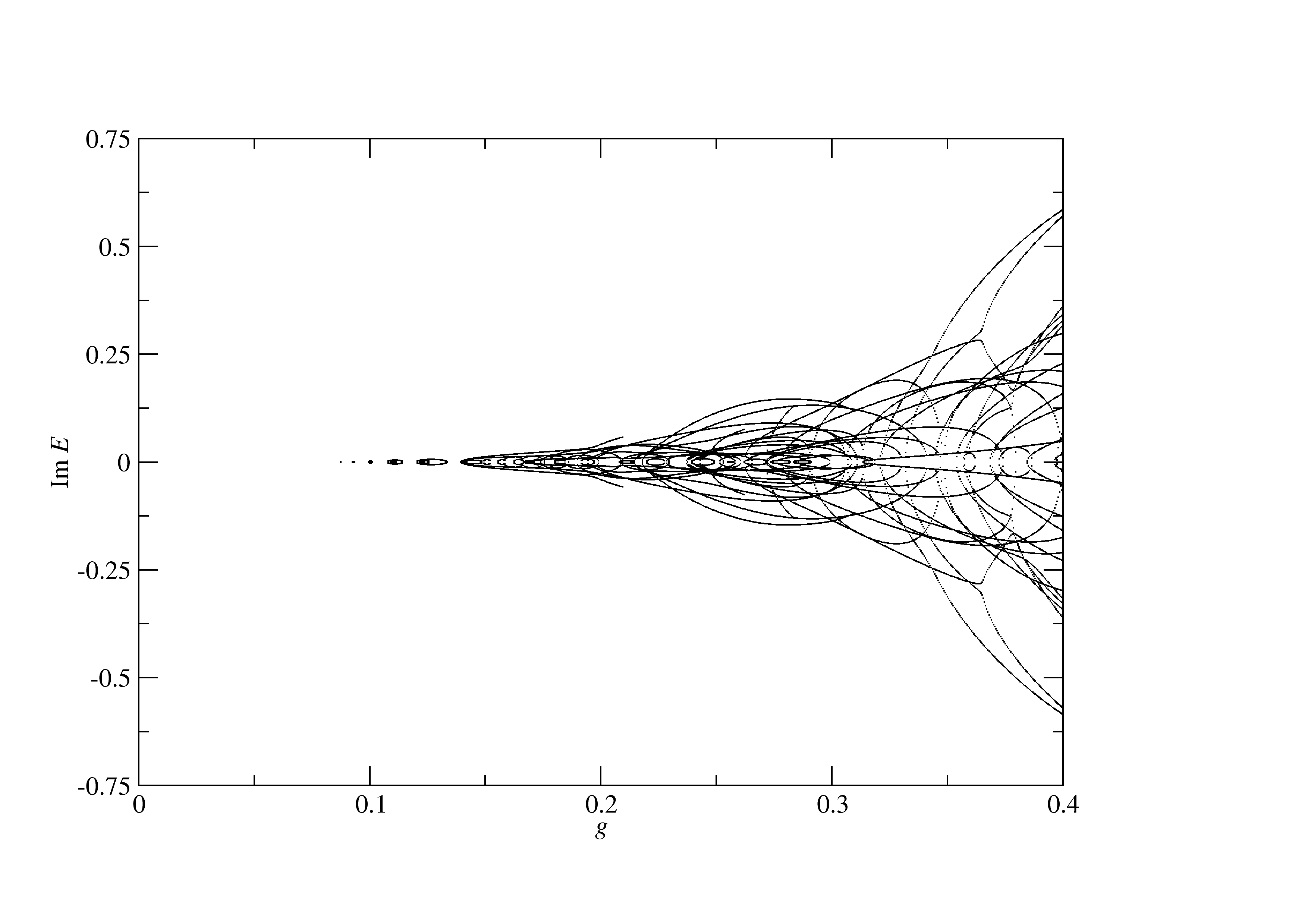}
\end{center}
\caption{Imaginary parts of the eigenvalues whose real parts are shown in
Fig.~\ref{F6}. Note that the imaginary parts of the eigenvalues are smaller than
the real parts by a factor of roughly 10. For the size of the matrix studied, it
appears that the critical point is near $g\approx 0.08$. Below this point we see
no complex eigenvalues whose real parts are less than 18. Further analysis
(see Fig.~\ref{F8}) suggests that the critical point is near $0.05$. }
\label{F7}
\end{figure}

For the Hamiltonian (\ref{E2}), we construct Figs.~\ref{F6} and \ref{F7}, which
are the analogs of Figs.~\ref{F3} and \ref{F4}. Note that the eigenvalues no
longer show a breaking of degeneracy as $g$ increases from $0$. In Fig.~\ref{F7}
we plot the imaginary parts of those eigenvalues. Note that there are no complex
eigenvalues when $g$ is below about $0.08$.

\begin{figure}[t!]
\begin{center}
\includegraphics[trim=10mm 17mm 53mm 28mm,clip=true,scale=0.50]{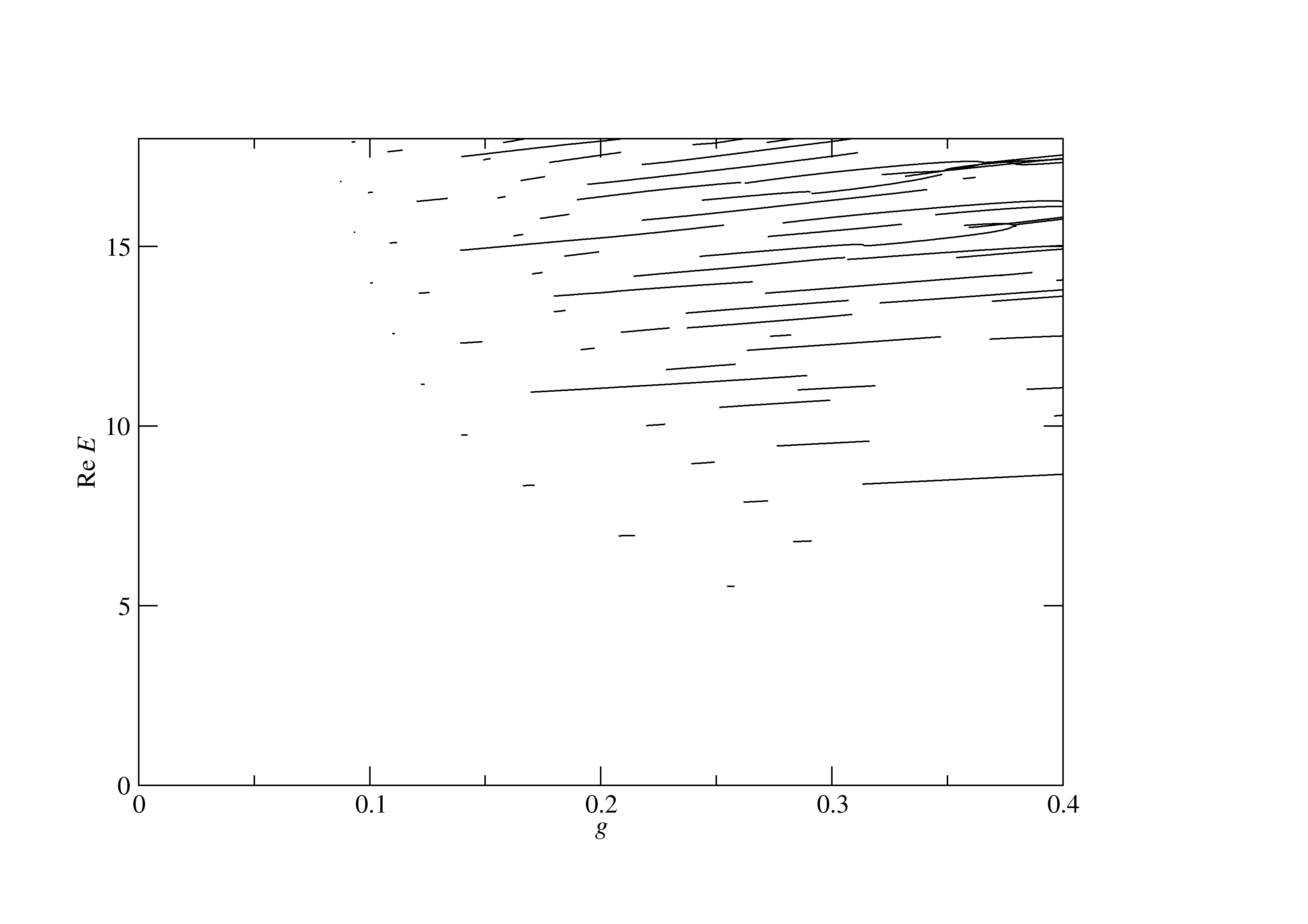}
\end{center}
\caption{Real parts of the eigenvalues in Fig.~\ref{F6} whose imaginary parts
are nonzero; the real parts of these eigenvalues grow with decreasing $g$. We
can obtain a more accurate determination of the critical value of $g$ by
fitting a curve through the left-most points on the graph, which are much
more regular than the left-most points in the analogous Fig.~\ref{F5}.}
\label{F8}
\end{figure}

From the eigenvalues in Fig.~\ref{F6} we select out just the eigenvalues that
have nonzero imaginary parts. The real parts of these eigenvalues are shown in
Fig.~\ref{F8}. As in the case of Fig.~\ref{F5}, these real parts increase
rapidly as $g$ decreases.

Because the left-most points in Fig.~\ref{F8} are quite regular, we can
numerically fit a curve that passes approximately through these points. If we
seek a curve of the form
\begin{equation}
f(g)=a(g-b)^c,
\label{E13}
\end{equation}
we find that
\begin{equation}
a=2.32\pm0.18,\quad b=0.046\pm0.002,\quad c=-0.615\pm0.033.
\label{E14}
\end{equation}
If we then try a more elaborate curve of the form
$$f(g)=a(g-b)^c[-\log(g-b)]^d,$$
we find that the value of $c$ is consistent with 0, which suggests an improved
fitting curve of the form
\begin{equation}
f(g)=a[-\log(g-b)]^d.
\label{E15}
\end{equation}
The numerical fit then gives
\begin{equation}
a=2.17\pm0.11,\quad b=0.054\pm0.001,\quad d=1.67\pm 0.5.
\label{E16}
\end{equation}
Thus, based on these extrapolations, we are moderately confident that there is a
critical value of $g$ near $0.05$, which is somewhat smaller than the rough
estimate of $g_{\rm crit}$ obtained by inspection of Fig.~\ref{F7}.

\subsection{Calculation of the Eigenvalues of the Hamiltonian (\ref{E3})}
\label{ss3c}

\begin{figure}[t!]
\begin{center}
\includegraphics[trim=10mm 17mm 53mm 28mm,clip=true,scale=0.50]{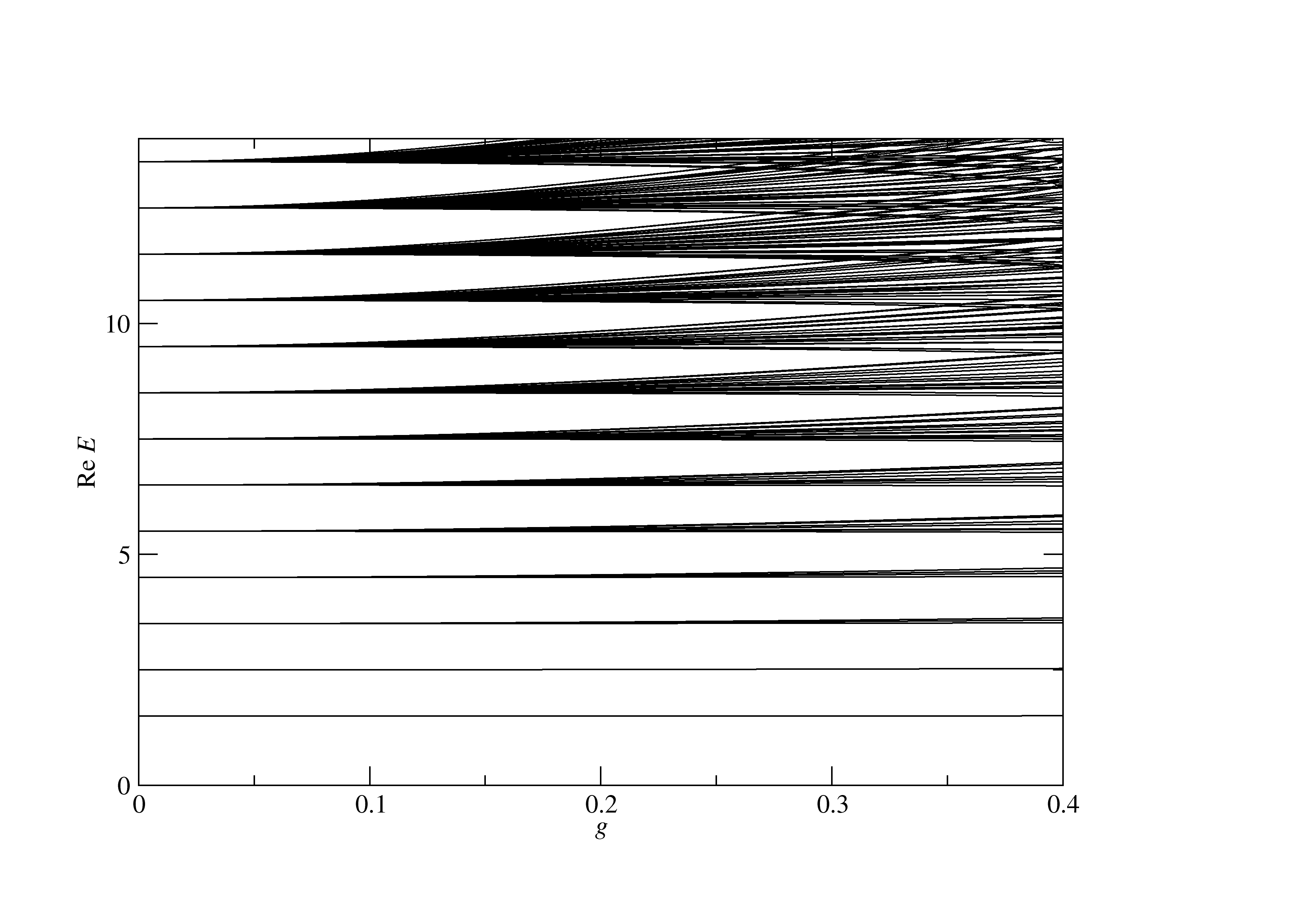}
\end{center}
\caption{Real parts of the eigenvalues of $H$ in (\ref{E3}) that have converged
for a $30^3\times 30^3$ matrix plotted versus $g$. The coupling constant $g$
ranges from $0$ to $0.4$ in steps of $0.0005$, and the eigenvalues shown have
real parts less than $14$. The plot is complicated because when $g=0$ the
eigenvalues become increasingly degenerate with increasing energy. There is one
eigenvalue of energy $1.5$, three of energy $2.5$, six of energy $3.5$, ten of
energy $4.5$, and so on. As $g$ increases, the degeneracy is broken. Eventually,
one can see energy levels crossing, but as in Fig.~\ref{F3}, most crossings only
represent accidental degeneracies. It is very difficult to find the exceptional
points.}
\label{F9}
\end{figure}

\begin{figure}[t!]
\begin{center}
\includegraphics[trim=10mm 17mm 53mm 28mm,clip=true,scale=0.50]{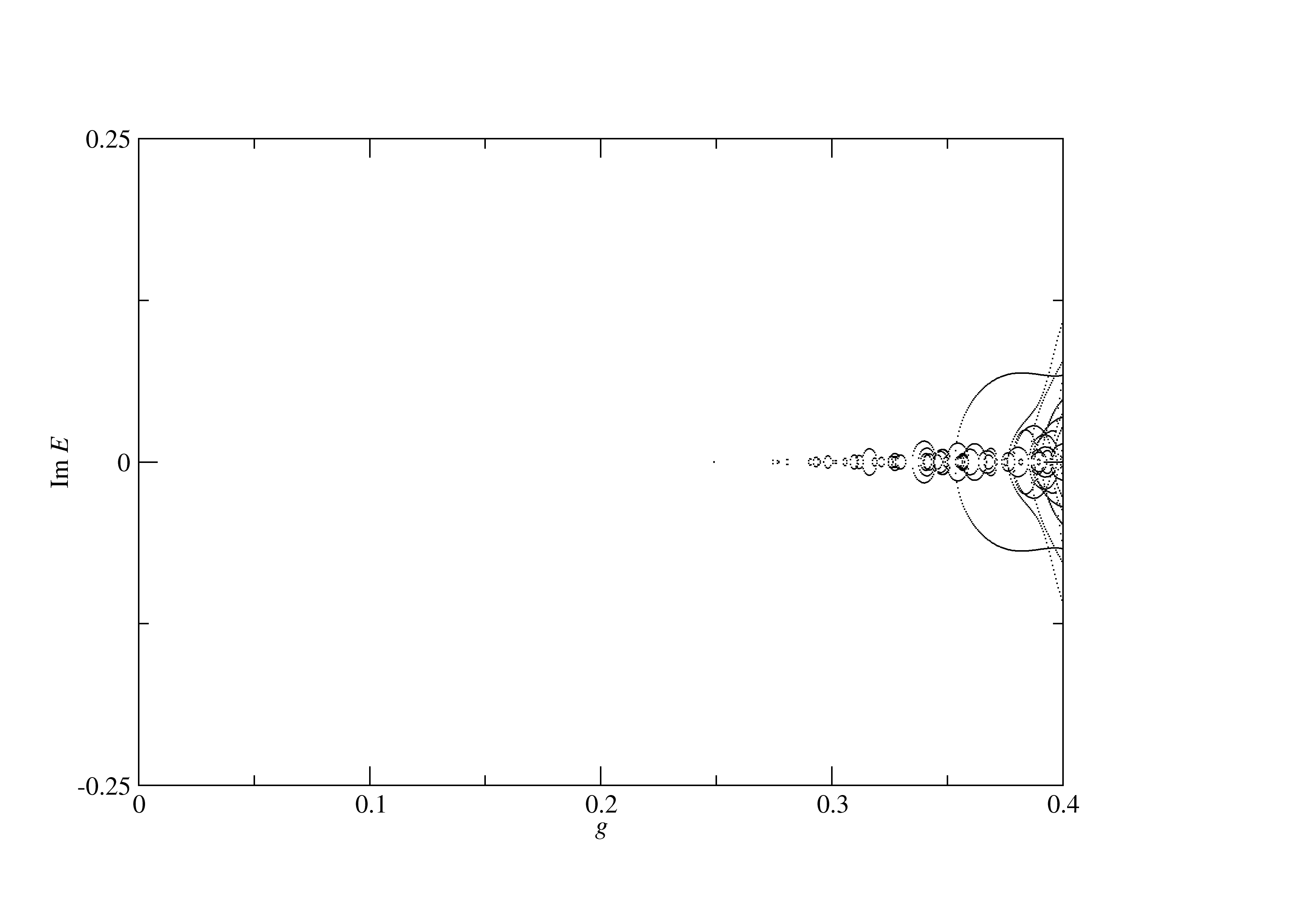}
\end{center}
\caption{Imaginary parts of the eigenvalues whose real parts are shown in
Fig.~\ref{F9}. There appears to be a critical point at $g\approx 0.25$. Below
this point there are no complex eigenvalues at all. The eigenvalue behavior
shown in this figure is similar to that shown in Figs.~\ref{F4} and \ref{F7} for
the Hamiltonians in (\ref{E1}) and (\ref{E2}).}
\label{F10}
\end{figure}

\begin{figure}[t!]
\begin{center}
\includegraphics[trim=10mm 17mm 53mm 28mm,clip=true,scale=0.50]{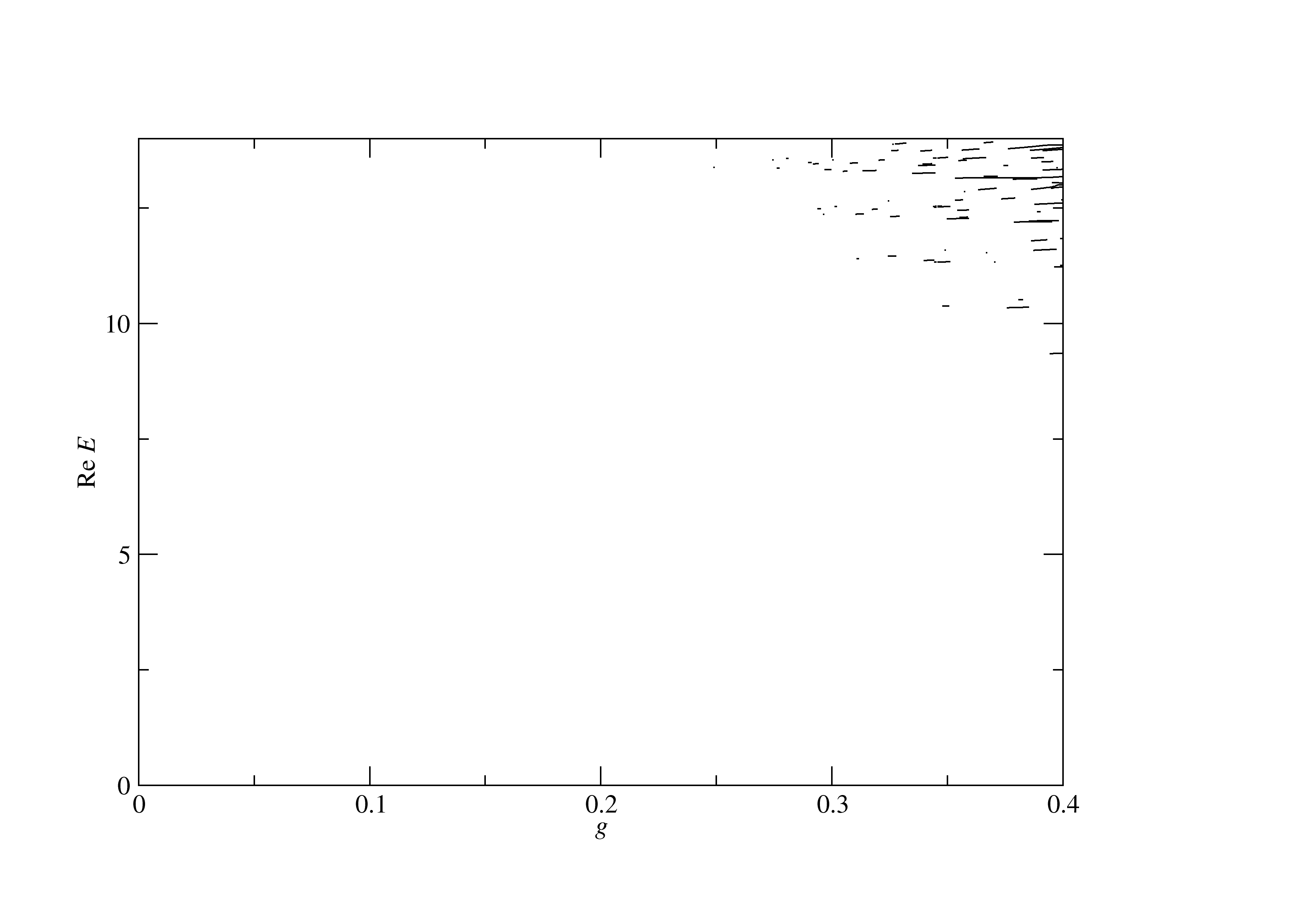}
\end{center}
\caption{Real parts of the eigenvalues in Fig.~\ref{F9} whose imaginary parts
are nonzero. Note that the real parts of these eigenvalues grow with decreasing
$g$. We can try to obtain a more accurate determination of the critical value of
$g$ by tracing a curve through the left-most points on the graph and estimating
that this curve blows up near $g\approx 0.2$. Unfortunately these points are
not regular enough to do a numerical fit to such a curve.}
\label{F11}
\end{figure}

Next we consider the Hamiltonian $H$ in (\ref{E3}). We diagonalize a $30^3\times
30^3$ matrix representation of $H$ and plot the results in Figs.~\ref{F9} and
\ref{F10}. Figure \ref{F10} suggests that there is a $\cP\cT$ phase transition
near $g\approx0.25$. These eigenvalues have stopped changing as the size of
the matrix size increases from $20^3\times 20^3$ to $25^3\times 25^3$ to
$30^3\times 30^3$. Then, in Fig.~\ref{F11} we isolate just those eigenvalues
whose nonzero imaginary parts are shown in Fig.~\ref{F10}.

\subsection{Calculation of the Eigenvalues of the Hamiltonian (\ref{E4})}
\label{ss3d}

\begin{figure}[t!]
\begin{center}
\includegraphics[trim=10mm 17mm 53mm 28mm,clip=true,scale=0.50]{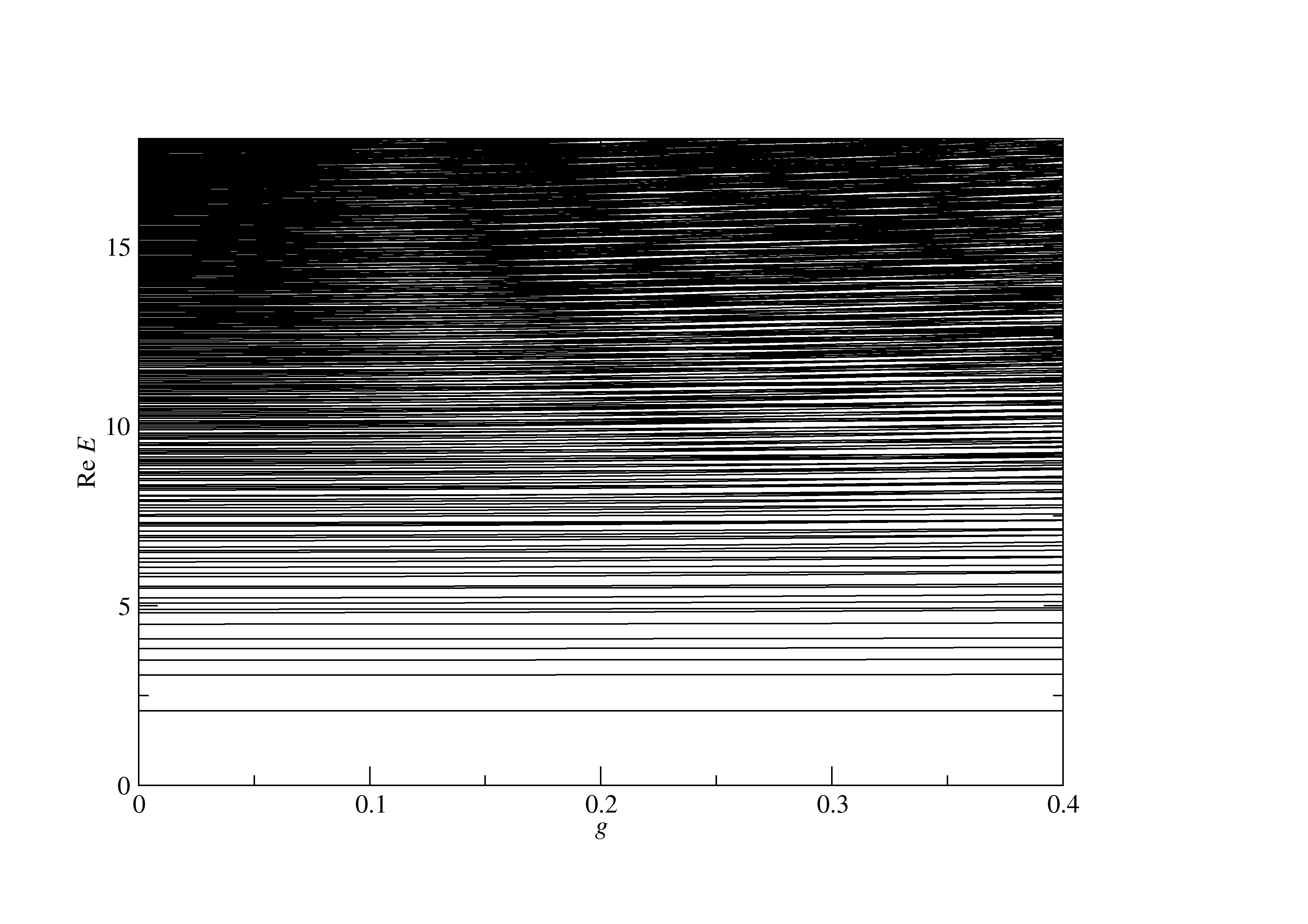}
\end{center}
\caption{Real parts of the eigenvalues of the $\cP\cT$-symmetric Hamiltonian
(\ref{E4}) plotted as functions of $g$ for $g$ ranging from $0$ to $0.4$ in
steps of $0.0005$. All eigenvalues (both real and complex) whose real parts are
less than 18 are shown.}
\label{F12}
\end{figure}

\begin{figure}[t!]
\begin{center}
\includegraphics[trim=10mm 17mm 53mm 28mm,clip=true,scale=0.50]{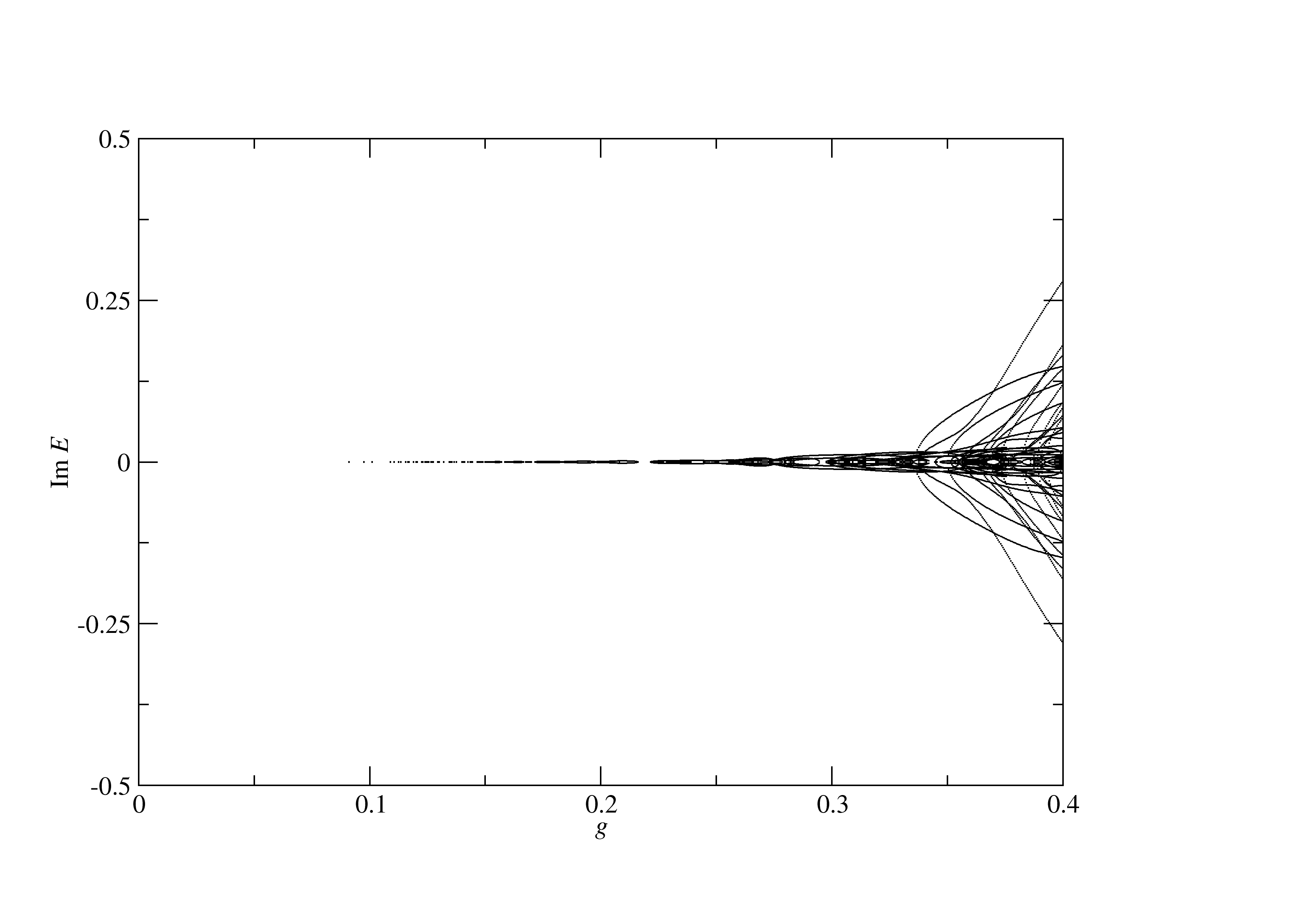}
\end{center}
\caption{Imaginary parts of the eigenvalues whose real parts are shown in
Fig.~\ref{F12}. There appears to be a critical point near $g\approx 0.09$. Below
this point there are no complex eigenvalues in the range studied. Further
analysis (see Fig.~\ref{F14}) suggests that the critical value of $g$ is near
$0.057$. The eigenvalue behavior shown in this figure is similar to that shown
in Figs.~\ref{F4}, \ref{F7}, and \ref{F10} for the Hamiltonians in
(\ref{E1}--\ref{E3}).}
\label{F13}
\end{figure}

\begin{figure}[t!]
\begin{center}
\includegraphics[trim=10mm 17mm 53mm 28mm,clip=true,scale=0.50]{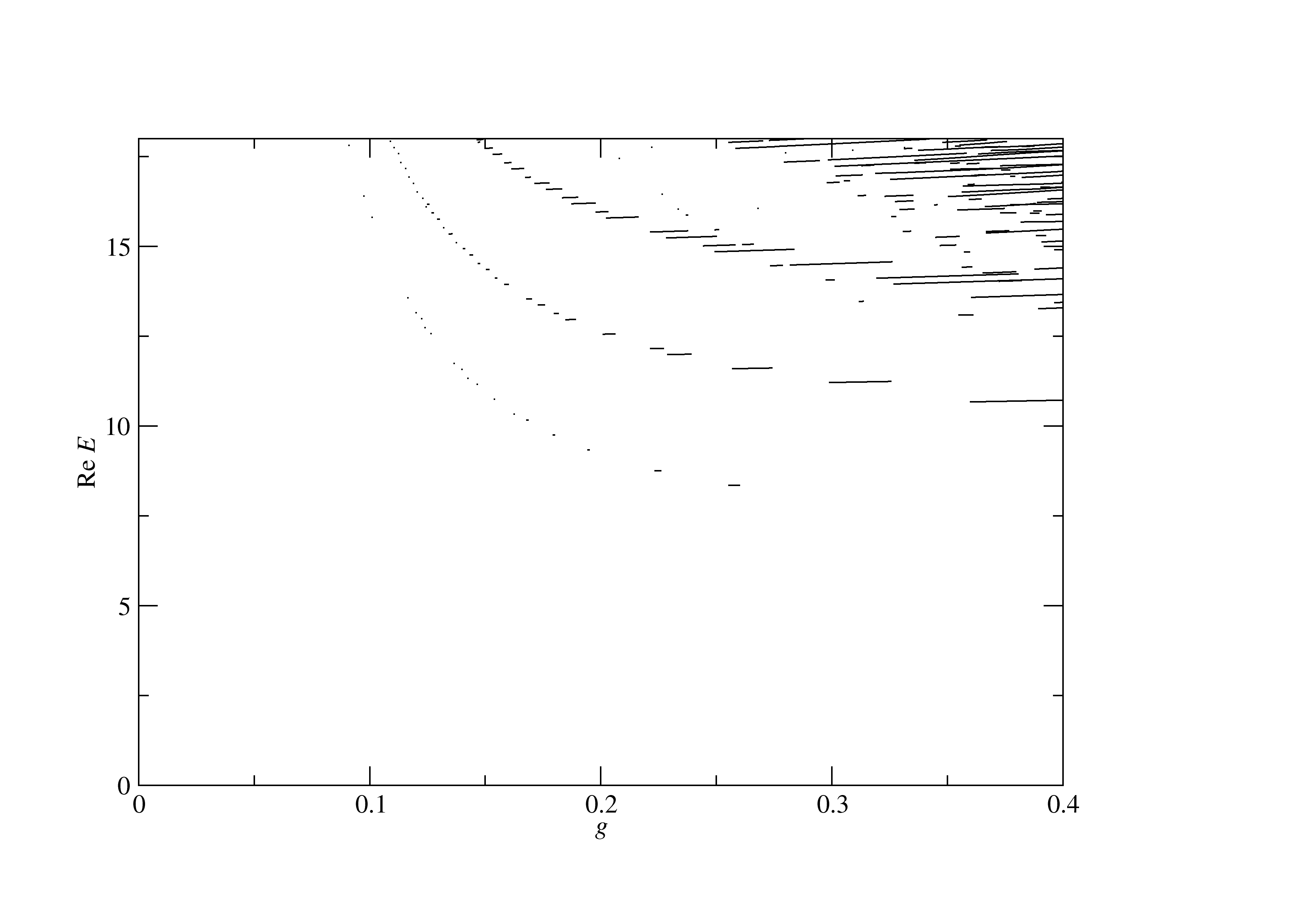}
\end{center}
\caption{Real parts of the eigenvalues in Fig.~\ref{F12} whose imaginary parts
are nonzero. These eigenvalues grow with decreasing $g$. A numerical fit to the
left-most points on the graph predicts that the critical value of $g$ is about
$0.05$.}
\label{F14}
\end{figure}

The analogs of Figs.~\ref{F9} and \ref{F10} for the Hamiltonian (\ref{E4}) are
Figs.~\ref{F12} and \ref{F13}. However, because the eigenvalue degeneracies have
been lifted, the left-most points in Fig.~\ref{F14} are regular enough to
perform fits of the form in (\ref{E13}) and (\ref{E15}). The first fit gives the
value $0.049\pm0.001$ for the critical point and the second fit gives a value of
$0.064\pm0.001$ for the critical point. Thus, we may estimate that there is a
critical value of $g$ near $0.057$, which is somewhat smaller than what one
would guess from Fig.~\ref{F13}.

\section{Summary}
\label{s4}

We have presented extensive numerical evidence in this paper that suggests that
the eigenvalues of the multidimensional trilinear Hamiltonians
(\ref{E1}--\ref{E4}) are entirely real when the coupling constant $g$ is less
than a critical value. If this numerical evidence stands up to further study,
then we can conclude that these quantum theories possess an unbroken $\cP
\cT$-symmetric phase for sufficiently small $g$ and display a transition to a
broken phase as $g$ increases. This would suggest that multidimensional $\cP
\cT$-symmetric quantum systems can exhibit the same kind of phase transition
that one-dimensional quantum systems are known to exhibit. On the basis of this
work we are tempted to conjecture that even $\cP\cT$-symmetric quantum field
theories such as an $ig\phi^3$ theory might exhibit a $\cP\cT$ phase transition
as the coupling constant $g$ increases from $0$.

In this paper we have only considered cubic and trilinear interactions because 
the harmonic-oscillator-basis functions that we have used to construct the large
numerical matrices whose eigenvalues we have obtained numerically have
exponentially vanishing asymptotic behaviors on the real axis. It would be
interesting to study quartic Hamiltonians using similar techniques, but we
anticipate that the numerical analysis would be much more difficult because the 
harmonic-oscillator-basis functions and the exact eigenfunctions have very
different asymptotic behaviors on the real axis.

Finally, we mention again that if there really is a $\cP\cT$-symmetric phase
transition for the models that we have studied, this transition is generically a
high-energy phenomenon. Unfortunately, this presents a severe problem for the
numerical techniques used in this paper because the Arnoldi method can only give
detailed information about converged eigenvalues for low energies as the size of
the matrix is increased. Thus, the principal finding in this paper, namely, that
the models we have studied exhibit a $\cP\cT$ phase transition, can only be
conjectural in nature.

\begin{acknowledgments}
CMB is supported by the U.K.~Leverhulme Foundation and by the U.S.~Department of
Energy. DJW thanks the Imperial College High Performance Computing Service for
the use of its resources.
\end{acknowledgments}

\end{document}